\documentclass[twoside,twocolumn,9pt]{article}
\usepackage{extsizes}
\usepackage[super,sort&compress,comma]{natbib} 
\usepackage[version=3]{mhchem}
\usepackage[left=1.5cm, right=1.5cm, top=1.785cm, bottom=2.0cm]{geometry}
\usepackage{balance}
\usepackage{mathptmx}
\usepackage{amsmath}
\usepackage{amssymb}
\usepackage{sectsty}
\usepackage{graphicx} 
\usepackage{lastpage}
\usepackage[format=plain,justification=justified,singlelinecheck=false,font={stretch=1.125,small,sf},labelfont=bf,labelsep=space]{caption}
\usepackage{float}
\usepackage{fancyhdr}
\usepackage{fnpos}
\usepackage[english]{babel}
\addto{\captionsenglish}{%
  
}
\usepackage{array}
\usepackage{droidsans}
\usepackage{charter}
\usepackage[T1]{fontenc}
\usepackage[usenames,dvipsnames]{xcolor}
\usepackage{setspace}
\usepackage[compact]{titlesec}
\usepackage{hyperref}

\usepackage{siunitx}
\DeclareSIUnit{\molar}{M}
\usepackage{tabularx}

\definecolor{cream}{RGB}{222,217,201}

\begin{document}

\pagestyle{fancy}
\thispagestyle{plain}
\fancypagestyle{plain}{
\renewcommand{\headrulewidth}{0pt}
}

\makeFNbottom
\makeatletter
\renewcommand\LARGE{\@setfontsize\LARGE{15pt}{17}}
\renewcommand\Large{\@setfontsize\Large{12pt}{14}}
\renewcommand\large{\@setfontsize\large{10pt}{12}}
\renewcommand\footnotesize{\@setfontsize\footnotesize{7pt}{10}}
\makeatother

\renewcommand{\thefootnote}{\fnsymbol{footnote}}
\renewcommand\footnoterule{\vspace*{1pt}%
\color{cream}\hrule width 3.5in height 0.4pt \color{black}\vspace*{5pt}} 
\setcounter{secnumdepth}{5}

\makeatletter 
\renewcommand\@biblabel[1]{#1}            
\renewcommand\@makefntext[1]%
{\noindent\makebox[0pt][r]{\@thefnmark\,}#1}
\makeatother 
\renewcommand{\figurename}{\small{Fig.}~}
\sectionfont{\sffamily\Large}
\subsectionfont{\normalsize}
\subsubsectionfont{\bf}
\setstretch{1.125} 
\setlength{\skip\footins}{0.8cm}
\setlength{\footnotesep}{0.25cm}
\setlength{\jot}{10pt}
\titlespacing*{\section}{0pt}{4pt}{4pt}
\titlespacing*{\subsection}{0pt}{15pt}{1pt}

\fancyfoot{}
\fancyfoot[LO,RE]{\vspace{-7.1pt}\includegraphics[height=9pt]{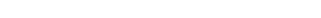}}
\fancyfoot[CO]{\vspace{-7.1pt}\hspace{13.2cm}\includegraphics{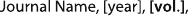}}
\fancyfoot[CE]{\vspace{-7.2pt}\hspace{-14.2cm}\includegraphics{head_foot/RF}}
\fancyfoot[RO]{\footnotesize{\sffamily{1--\pageref{LastPage} ~\textbar  \hspace{2pt}\thepage}}}
\fancyfoot[LE]{\footnotesize{\sffamily{\thepage~\textbar\hspace{3.45cm} 1--\pageref{LastPage}}}}
\fancyhead{}
\renewcommand{\headrulewidth}{0pt} 
\renewcommand{\footrulewidth}{0pt}
\setlength{\arrayrulewidth}{1pt}
\setlength{\columnsep}{6.5mm}
\setlength\bibsep{1pt}

\makeatletter 
\newlength{\figrulesep} 
\setlength{\figrulesep}{0.5\textfloatsep} 

\newcommand{\topfigrule}{\vspace*{-1pt}%
\noindent{\color{cream}\rule[-\figrulesep]{\columnwidth}{1.5pt}} }

\newcommand{\botfigrule}{\vspace*{-2pt}%
\noindent{\color{cream}\rule[\figrulesep]{\columnwidth}{1.5pt}} }

\newcommand{\dblfigrule}{\vspace*{-1pt}%
\noindent{\color{cream}\rule[-\figrulesep]{\textwidth}{1.5pt}} }

\makeatother

\twocolumn[
  \begin{@twocolumnfalse}
{\includegraphics[height=30pt]{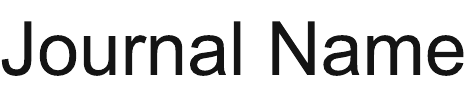}\hfill\raisebox{0pt}[0pt][0pt]{\includegraphics[height=55pt]{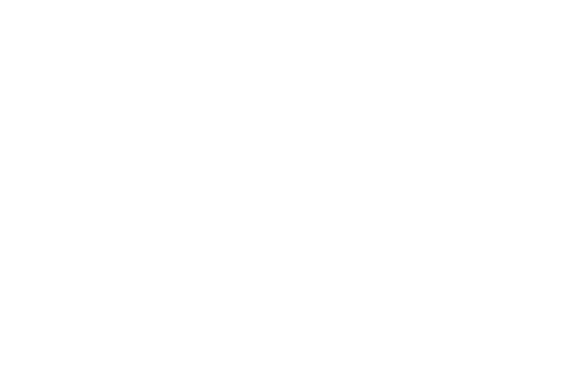}}\\[1ex]
\includegraphics[width=18.5cm]{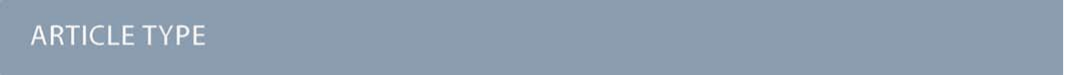}}\par
\vspace{1em}
\sffamily
\begin{tabular}{m{4.5cm} p{13.5cm} }

\includegraphics{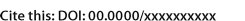} & \noindent\LARGE{\textbf{%
Mechanical response of a simple DNA nanostar hydrogel: symptoms of disorder and glassy emergence of solidity.%
$^\dag$}} \\
\vspace{0.3cm} & \vspace{0.3cm} \\

 & \noindent\large{Hajar Ajiyel,\textit{$^{a,b,d}$} Anthony J. Genot,$^\ddag$\textit{$^{b}$} Soo Hyeon Kim,\textit{$^{b}$} Nicolas Schabanel,\textit{$^{c}$} Hervé Guillou,\textit{$^{a}$} Catherine Barentin,\textit{$^{d}$} and Mathieu Leocmach$^\ast$\textit{$^{d}$}} \\

\includegraphics{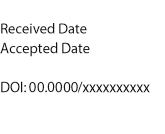} & \noindent\normalsize{
    DNA self-assembly is a well-understood nanotechnology to obtain extremely ordered structures from the nanometer to up to the hundred of microns scale. By contrast, DNA hydrogels rely on the disordered assembly of DNA building blocks to reach macroscopic volumes. However, in order to hold the promise of DNA bulk materials, the sequence designer needs a systematic understanding of how macroscopic properties emerge from disorder. 
    Here, we show a method to study systematically the mechanical response of a simple DNA nanostar hydrogel. This method mobilises bulk rheology, dynamic light scattering microrheology, mechanical modeling, as well as thermodynamic calculation and DNA sequence alteration.
    At low temperatures, we demonstrate a systematic deviation from Maxwell behaviour that is symptomatic of disordered materials. At temperatures much higher than the percolation of the DNA network, we characterise a surprising solid behaviour that we attribute to a glass transition. 
    Our results show the importance of disorder in DNA materials. Furthermore, the method we showcase in this article can be widely applied to more complex DNA materials. 
} \\

\end{tabular}

 \end{@twocolumnfalse} \vspace{0.6cm}

  ]

\renewcommand*\rmdefault{bch}\normalfont\upshape
\rmfamily
\section*{}
\vspace{-1cm}


\footnotetext{\textit{$^{a}$~Institut Néel-CNRS/UGA-Grenoble INP, Grenoble, France.}}
\footnotetext{\textit{$^{b}$~LIMMS-CNRS/Institute of Industrial Science-The University of Tokyo, Tokyo, Japan.}}
\footnotetext{\textit{$^{c}$~LIP-CNRS/ENS Lyon, Lyon, France.}}
\footnotetext{\textit{$^{d}$~Universite Claude Bernard Lyon 1, CNRS, Institut Lumière Matière, UMR5306, F-69100, Villeurbanne, France.}}

\footnotetext{\ddag~Anthony Genot passed away on April 8th 2025.}
\footnotetext{$\ast$~E-mail: mathieu.leocmach@univ-lyon1.fr.}



\section{Introduction}

In recent years, DNA has emerged as an ideal polymer to build soft materials. In addition to being mechanically robust, chemically stable and enzymatically replicable, DNA is a sequence-defined polymer.
Two single-stranded DNA of complementary sequences can hybridize into the canonical double helix, forming a rigid bond.
By smartly programming the monomer sequence, DNA can self-assemble into almost any shape~\cite{Rothemund2006}.
DNA nanostructures have been designed to dynamically respond to their chemical or mechanical environment by changing shape, disassembling, emitting light, etc.~\cite{Goodman2008, Merindol2019}.
Structures of hundreds of nanometers and hundreds of strands of complex shapes called `DNA origami' are now routinely assembled (provided sufficient charge screening) from computer-aided design~\cite{Zadeh2011, levy_et_al:LIPIcs.DNA.27.5} relying on the thermodynamics of DNA assembly to encode a geometry.
Larger ordered structures can be produced~\cite{zheng2009MolecularMacroscopicRational}, up to hundreds of microns. However, such precise assemblies do not scale up well to bulk materials. To obtain centimeter-scale DNA materials, one has to embrace disorder and rely on the random assembly of DNA building blocks, sometimes associated to peptides, nanoparticles, synthetic polymers, etc. These building blocks form a space-spanning network called a DNA hydrogel~\cite{Li2019,Wang2017}.
Due to the disordered nature of DNA hydrogels, predicting their mechanical properties from the DNA sequence is a challenge.

In the present article, we focus on the simplest DNA hydrogel, and show how its mechanical properties can be characterized and understood down to the base pair level using methods from rheology, soft matter and statistical physics. In particular we demonstrate the fundamental role disorder at the nanoscale plays in the mechanical properties at the macroscale.

\begin{figure}
    \centering
    \includegraphics[width=\linewidth]{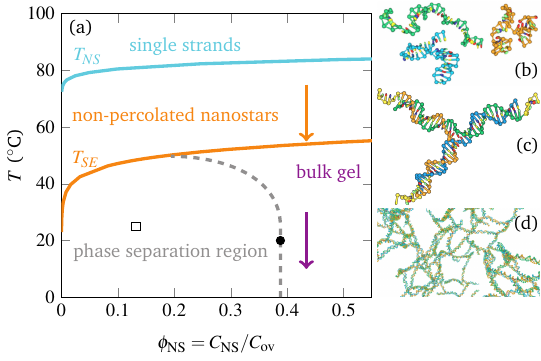}
    \caption{(a) Theoretical phase diagram of DNA nanostars \texttt{Y16SE6}. The cyan and orange lines are the concentration dependence of the melting temperatures of the nanostars without sticky ends $T_\mathrm{NS}$, respectively the duplex of sticky ends $T_\mathrm{SE}$. The black dot is experimentally determined gel concentration at coexistence. The gray dashed line is a guide for the eye. Orange and purple arrows show concentration and the respective temperature ranges of DLS microrheology and bulk rheology measurements.
    (b) Snapshot of the configuration of the three isolated DNA strands at high temperature from \texttt{oxDNA} simulations. 
    (c) Same for an isolated nanostar at intermediate temperature.
    (d) Same for a gel at the statepoint shown as an empty square in (a).}
    \label{fig:phase-diag}
\end{figure}

Our system belongs to the well-studied class of DNA nanostar hydrogels~\cite{Biffi2013, Rovigatti2014-2, Biffi2015, Nava2017,Fernandez2018, Conrad2019, Sato2020, Conrad2022, chaderjian2026DiverseDistinctDensely}. It is composed solely of 3 strands of DNA that self-assemble into a 3-branch motif called a nanostar. Each arm ends in a palindromic ssDNA overhang, called a sticky end, that can bind the arms of two different nanostars.
The assembly of such DNA nanostar hydrogel is often described in two steps shown in Fig.~\ref{fig:phase-diag}.
At high temperatures, all DNA is fully denatured with isolated ssDNA strands. Below a temperature called $T_\mathrm{NS}$ the nanostars self-assemble. At a lower temperature $T_\mathrm{SE}$ the sticky ends hybridize with one another linking neighbouring nanostars. At low temperature, we get a hydrogel network.



The phase behavior of the nanostar system depends not only on the temperature but also on the initial concentration of DNA, as shown on Fig.~\ref{fig:phase-diag}. At high enough concentration and below $T_\mathrm{SE}$, the network of nanostars spans space, whereas at low concentrations a dense phase rich in DNA coexists with a dilute phase poor in DNA~\cite{Biffi2013, Biffi2015, Conrad2019, Conrad2022, Rovigatti2014-2}.
It has been shown that the position of the phase boundaries is influenced by the number of arms (valency) of the nanostars~\cite{Biffi2013, Biffi2015} and the length of the sticky ends~\cite{Sato2020}.
The phase separation pattern and kinetics are strongly altered by minute DNA sequence changes that affect the flexibility of the design~\cite{Nguyen2017}.
Even though the self-assembly of the monomer motif is very precise and organized, the hydrogel network is intrinsically disordered~\cite{Rovigatti2014-2,Locatelli2017}. Furthermore, at high enough temperatures, the lifetime of sticky-end bonds is short~\cite{Nava2017,Biffi2015,Conrad2019} and not all sticky ends are hybridized simultaneously. We thus have a disordered temperature-controlled dynamic network, the perfect test bed for a statistical physics approach to prediction of mechanical properties.


The main framework used to understand the statistical physics of DNA nanostar hydrogels is the patchy particles approach; where the nanostars are modeled as rigid particles with as many attractive patches as arms. The patches model the sticky-end interactions as they are selective, tunable, and directional. Through this modeling, the hydrogel formation is found to be similar to strong glasses \cite{Biffi2013,Biffi2015}, meaning that the timescales follow an Arrhenius trend. This relation is parametrized by the enthalpy of hybridization of the sticky ends~\cite{Nava2017,Biffi2015,Conrad2019}. This results in very large variations in relaxation timescales of the material, depending on temperature. 
This model, together with percolation theory of limited-valence network, helped to understand the non-Newtonian response of DNA nanostar hydrogels at temperatures below $T_\mathrm{SE}$~\cite{Nava2018}. Alternatively, bulk rheological measurements below $T_\mathrm{SE}$ have been modeled using Maxwell isostaticity~\cite{Conrad2019} or topology~\cite{Palombo2025}.
Above $T_\mathrm{SE}$, the mechanical response of advanced DNA nanostar hydrogel designs (intermediate linker~\cite{Xing2018}, competing strands~\cite{Fernandez2018, bomboi2019ColdswappableDNAGels}) have been characterized by dynamic light scattering (DLS) microrheology, with the simplest design treated as a mere point of reference. Indeed, above $T_\mathrm{SE}$, the bonded sticky-ends are a minority, the network of nanostars does not span space and a simple fluid response is expected.

In this study, we will focus on the simplest DNA nanostar design and walk through the methodology of its investigation by the tools of soft matter and statistical physics.
First, we will investigate its phase diagram and identify in which region the mechanical behavior is expected to be well-defined and informative. In our simple case, it means restricting our study to a large concentration regime where only continuous phases are expected.
Then, we will estimate the variation of the characteristic time scale of the material depending on temperature, here a 7 orders of magnitude range. We will explain how to select complementary techniques to cover this range of time scales. In our case, we select bulk rheology and DLS microrheology and explain their respective principles.
Next, we will show how a careful comparison between mechanical measurements and models can yield an understanding of nanoscale structure and dynamics. In particular, we will show that the `simplest DNA nanostar design' becomes solid at much higher temperatures than expected. This hints towards a solidification mechanism that does not require percolation of the network.
Finally, we will draw on the versatility of DNA technology to confront our hypothesis. We will alter the DNA nanostar design in subtle and controlled ways, measure the resulting mechanical response using the same methodology, and confirm the consistency of our model.
Through this methodical exploration, we will show that DNA nanostars at high concentrations solidify in two steps upon cooling: a glass transition, followed by gelation.




\section{Materials and methods}
\subsection{A quick tutorial about linear mechanics}\label{sec:linearmech}
At its simplest, mechanics is a study of the response of a material to a stress (force applied on an area) in terms of strain (ratio of the final size by the initial size), or the reverse. For instance, one can suddenly apply a constant stress $\sigma$ and measure the resulting time-dependent strain $\gamma(t)$. The ratio $J(t) = \gamma(t)/\sigma$ is called the creep compliance. From the study of material response functions like $J(t)$, one can extract materials properties like characteristic times (in \unit{\second}), elastic moduli (in \unit{\pascal}) or viscosities (in \unit{\pascal\second}). In turn, material properties can be related to microscopic quantities like bond lifetime or the mean distance between nodes in a network. Furthermore, the temperature dependence of material properties can be linked to thermodynamic quantities (e.g. enthalpy and entropy of binding).

All these interpretations rely on models at two levels. 
(i) Extracting microscopic quantities from material quantities rely on hypothesis about the microstructure, which are easier to come by and to test when the microstructure has been designed and can be altered. That is where the versatility of DNA shines. 
(ii) Extracting material quantities from mechanical response functions rely on mechanical models and the theory of response. 
For instance, a purely elastic solid can be modelled by a spring that responds to a step stress as $J(t)=1/G$ where $G$ is the elastic modulus. A purely viscous fluid can be modelled by a dashpot that responds as $J(t)=t/\eta$ where $\eta$ is the viscosity. Many soft materials are instead modelled by a so-called `spring-pot' that responds as $J(t)=t^\alpha/\mathbb{V}$, where $\alpha$ is an exponent that varies between 0 (solid) and 1 (liquid) and $\mathbb{V}$ (in \unit{\pascal\,\second^\alpha}) is a quasi-property that can be considered as the modulus at $t=\qty{1}{\second}$ if $\alpha<0.5$ or the viscosity at $t=\qty{1}{\second}$ if $\alpha>0.5$~\cite{Jaishankar2013}. 
Materials that have different responses depending on the time scale are modelled by a combination of springs, dashpots and spring-pots either in series or in parallel. The response functions of such models to step stress, oscillatory strain, or other mechanical stimuli can be deduced analytically (or at least numerically for more involved models), which allow fitting their parameters on experimental responses to deduce material properties.

Dynamic networks are often described by a Maxwell model, reducing the complexity of the material to a spring in series with a dashpot, quantified respectively by an elastic modulus $G$ and a viscosity $\eta$. At time scales shorter than the characteristic time $\tau=\eta/G$ the Maxwell model behaves elastically like a solid, whereas at longer time scales it behaves like a viscous fluid. 

\subsection{System of study}

Following~\cite{Sato2020}, we study mainly a 3-arm DNA nanostar (Y motif, shown in Fig.~\ref{fig:phase-diag}c), each arm composed of a 16~bp (base pairs) double-stranded segment immediately followed by a 6~bp palindromic overhang called a sticky end (SE, yellow on Fig.~\ref{fig:phase-diag}c) that is identical for each arm. The SE junction has thus little flexibility. By contrast, the junction between the arms at the center of the nanostar is flexible due to two consecutive unpaired thymine on each strand. In the following we call this design \texttt{Y16SE6}. We will also compare it to \texttt{Y16SE0} (no sticky-end), \texttt{Y16SE4} (4~bp sticky-end), \texttt{Y16SE8} (8~bp sticky-end), or \texttt{Y32SE6} (32~bp arms). We tuned the sequence design of the nanostar with 32~bp arms (see Table~\ref{tab:Y32}) so that the melting temperature is within \qty{1}{\kelvin} of the nanostar with 16~bp arms.
DNA sequences for the nanostars with 16~bp arms can be found in Ref\cite{Sato2020}.

\begin{table}[b]
    \begin{tabularx}{\columnwidth}{lX}
         \# & Sequence 5'-3'\\
        1 & \texttt{GCTAGC}-\texttt{\textcolor{blue}{ACAATACCCGGTTTGCATGGTTAAGACGTAAT}}-\texttt{TT}-\linebreak\texttt{\textcolor{red}{TTAATATTCAGTGTATCATCCATATTCGCCAA}}\\
        2 & \texttt{GCTAGC}-\texttt{\textcolor{red}{TTGGCGAATATGGATGATACACTGAATATTAA}}-\texttt{TT}-\linebreak\texttt{\textcolor{green!60!black}{GATTTCAATAATGTTTCAAGGAGATAGTTGGA}}\\
        3 & \texttt{GCTAGC}-\texttt{\textcolor{green!60!black}{TCCAACTATCTCCTTGAAACATTATTGAAATC}}-\texttt{TT}-\linebreak\texttt{\textcolor{blue}{ATTACGTCTTAACCATGCAAACCGGGTATTGT}}
    \end{tabularx}
    \caption{Sequence of the three strands of \texttt{Y32SE6}, each 72~bp long.}
    \label{tab:Y32}
\end{table}

DNA sequences are ordered from IDT and Eurogentec in dried form, with only a desalted purification.
During DNA oligomer synthesis, nucleotides are added one by one. There is a probability $q$ that a nucleotide does not attach on any given strand. If so, the strand is capped and will not grow further. Although there is also a small failure rate of capping, we will not take it into account in the following. According to our suppliers, $q\approx\num{e-2}$. During desalted purification, strands shorter than 10 bases are eliminated. Given $N$ the desired length, the probability that a strand reaches length $n$ is thus
\begin{align}
P(n) &= (1-q)^{N-10}\text{, for }n=N\label{eq:proba_full_length}\\
&= (1-q)^{n-10} q \text{, for }n\in[10,N).\label{eq:proba_length_n}
\end{align}

Strands are synthesised with the sticky-end coming last, therefore we can split strands in three populations depending on how early the termination took place : (i) perfect strands, (ii) strands that lack part of their sticky-end but are fully able to assemble into a nanostar, and (iii) strands where nanostar assembly is defective. For \texttt{Y16SE6} each population amounts respectively to 74\%, 4.6\% and 21.4\%, according to Eq.~(\ref{eq:proba_full_length}) and Eq.~(\ref{eq:proba_length_n}).

In a well-annealed situation, nanostar assembly is cooperative and strands with two full arms (with or without SE) will assemble more readily together. We will thus consider that the probability to assemble a nanostar from 3 strands from categories (i) or (ii) is $P(n\geq N-n_\mathrm{SE})$ (and not $P(n\geq N-n_\mathrm{SE})^3$ as in the hypothesis of non cooperative assembly).

We use calibrated micropipettes. Each strand is diluted to \qty{5}{\milli\molar}, stored in \qty{50}{\milli\molar} phosphate buffer \qty{0.1}{\milli\molar} EDTA. For use in measuring sample properties \qty{300}{\milli\molar} \ce{NaCl} is added to the final solution. The DNA concentration, i.e. the number of nucleotide bases in the system, is estimated by absorbance spectroscopy and converted to \emph{nominal} strand molar concentration by supposing that every strand is $N$ bases long. However the real concentration has to take into account the mean strand length $\bar{n} = \sum_{n=10}^{N} n P(n)$.
The ratio between actual fully-formed nanostar concentration and nominal nanostar concentration is thus
\begin{equation}
\frac{C_\mathrm{NS}^\mathrm{real}}{C_\mathrm{NS}} = P(n\geq N-n_\mathrm{SE}) \frac{N}{\bar{n}}.
\label{eq:real_C_NS}
\end{equation}
For \texttt{Y16SE6} this correction factor is $\approx 0.88$.%

In the following, we will identify samples by their nominal concentrations ($C_\mathrm{NS}=\qty{1}{\milli\molar}$ in all measured samples, unless mentioned otherwise). However Eq~(\ref{eq:real_C_NS}) has been used in all calculations to work with $C_\mathrm{NS}^\mathrm{real}$.
For all tests, the sample is well mixed and annealed at \qty{90}{\celsius} during at least \qty{5}{\minute}, then the temperature is lowered at \qty{1}{\celsius\per\min} to the measuring temperature. 

\subsection{Establishing the phase diagram}
Theoretical melting curve of nanostars without sticky end is obtained at any concentration $C_\mathrm{NS}$ via \texttt{Nupack}~\cite{Zadeh2011} calculation from the sequence of the three strands of \texttt{Y16SE0}, each at concentration $C_\mathrm{NS}$. We define the melting temperature of the nanostar $T_\mathrm{NS}(C_\mathrm{NS})$ as the temperature where half of the nanostars are assembled.
Neglecting the presence of the sticky ends may overestimate $T_\mathrm{NS}$ by $\approx \qty{2}{\kelvin}$~\cite{dimichele2014EffectInertTails, soto2025ModellingMeltingDNA}.%
In the same way, theoretical melting temperature of the SE-SE bond $T_\mathrm{SE}(C_\mathrm{NS})$ is determined by considering in \texttt{Nupack} a system made of the three possible doubled-stranded \qty{16}{bp} arms terminated by sticky ends. Any two arm can bind by their sticky ends.
This methods takes into account the stacking interaction between the $5^\prime$ terminal base of a sticky end and the $3^\prime$ terminal base of the strand of the nanostar arm. 
Since this $3^\prime$ terminal base is different between the arms, we observe that the bonding between different pairs of arms lead to different bonding probabilities and free energies. In the following we consider average values but keep in mind the underlying distributions. In particular the melting temperature is determined from the average SE-SE bond probability such that  $p_\mathrm{SE}(T_\mathrm{SE})=0.5$.
\texttt{Nupack} calculations are based on thermodynamic data from~\cite{SantaLucia1998}. The results are displayed on Fig~\ref{fig:phase-diag} as cyan and orange curves respectively.


The coexistence concentration was determined by preparing a sample at \qty{5}{\micro\molar} of each strand in a BSA-coated well, together with Evagreen for fluorescent marking. To facilitate the phase separation, the sample was incubated for \qty{1}{\hour} with continuous 450~rpm shaking at \qty{37}{\celsius}, a temperature just below $T_\mathrm{SE}$, followed by centrifugation (well-plate centrifuge 3600~rpm) until all the DNA dense phase was gathered in a single droplet ($\approx \qty{400}{\micro\metre}$ laterally) containing DNA-poor phase inclusions. Then we acquired a 3D stack of the droplet using an Olympus laser scanning confocal microscope at 20x magnification that we segment to deduce the volume of the dense phase. We derive a nominal coexistence concentration of $\qty{894\pm10}{\micro\molar}$ reported on Fig~\ref{fig:phase-diag} as a black dot. 

We estimate the contour radius of a nanostar as $R = n_\mathrm{Y}u_\mathrm{ds}+\sqrt{3}u_\mathrm{ss}$, where $n_\mathrm{Y}=16$ is the number of base pairs in the double-stranded segments, $u=\qty{0.332}{\nano\metre/\mathrm{bp}}$ the rise per base pair in B-DNA, and $u_\mathrm{ss}=\qty{0.43}{\nano\metre/\mathrm{bp}}$ the contour length per base pair of single-stranded DNA to take into account the central flexible part of the nanostar.
To convert from contour radius $R$ to effective length $R_\mathrm{eff}$, we have to take into account the persistence length of DNA $\ell_\mathrm{p}\approx\qty{50}{\nano\metre}$ in our high salt condition~\cite{baumann1997IonicEffectsElasticity}. Following~\cite{Dobrynin1995} we use the expression
\begin{equation}
    R_\mathrm{eff} = R(1+R/\ell_\mathrm{p})^{-2/5}
    \label{eq:crossoverSAW}
\end{equation}
to describe the crossover between the rigid rod regime and the self-avoiding walk. We obtain $R_\mathrm{eff}(n_\mathrm{Y}=16)\approx\qty{5.80}{\nano\metre}$.

The Brownian rotational time of an object of effective radius $R_\mathrm{eff}$ is $\tau_\mathrm{rot} = 8\pi\eta_\mathrm{s} R_\mathrm{eff}^3/(k_\mathrm{B}T)$, where $\eta_\mathrm{s}$ is the viscosity of the buffer of the order of \qty{1}{\milli\pascal\second}~\cite{doi1988theory}. For a single nanostar $\tau_\mathrm{rot,1} \approx\qty{1}{\micro\second}$. Therefore for all longer time scales, unbound nanostars can be considered as randomly rotating spheres of radius $R_\mathrm{eff}$.
The theoretical molar overlap concentration between nanostars is obtained as $C_\mathrm{ov}=\num{3e3}/(4\pi R_\mathrm{eff}^3\mathcal{N}_\mathrm{A})$, where $\mathcal{N}_\mathrm{A}$ is the Avogadro number. The resulting $C_\mathrm{ov}\approx\qty{2.03}{\milli\molar}$ is used to scale the concentration axis in Fig.~\ref{fig:phase-diag}.

In order to perform reliable mechanical measurements, we need continuous phases and thus to remain outside of the phase separation region. Therefore, we decided to work at a nominal concentration of $\qty{1}{\milli\molar}$ ($C_\mathrm{NS}^\mathrm{real}\approx\qty{0.88}{\milli\molar}$), that is both safely above coexistence and well below overlap.

\subsection{Time scale}
For a dynamic network as DNA nanostar hydrogels, the expected mechanical behavior is Maxwellian~\cite{Conrad2019}, that is solid-like at time scales much shorter than the bond lifetime, and liquid-like at much longer time scales. Measuring response functions at time scales well into both regimes is necessary to fully characterize the system.

Previous studies~\cite{Nava2017,Biffi2015,Conrad2019} have found that the rearrangement time scale of DNA nanostars follows
\begin{equation}
    \tau = \tau_0\exp\left(\frac{n(\Delta H - T\Delta S)}{RT}\right),
    \label{eq:tau}
\end{equation}
where $\Delta H$ and $\Delta S$ are the molar enthalpy and molar entropy of the SE-SE bond, $\tau_0$ is the timescale at $T=T_\mathrm{SE}$, and $n$ is the number of SE-SE bonds that need to break to enable rearrangement. It has been argued that $n$ should depend on the valency, e.g. $n=2$ in the case of 3-arm nanostars~\cite{Nava2017,Biffi2015}. However rheological studies have found an activation energy independent of the valency and corresponding to $n\approx 1$~\cite{Conrad2019}.
At our concentration, we estimate $\Delta H\approx\qty{211}{\kilo\joule/\mol}$ and $\Delta S\approx\qty{644}{\joule/\mol/\kelvin}$ according to our \texttt{Nupack} simulations.
We thus expect a 7 orders of magnitude ratio between timescales at \qty{90}{\celsius} and at \qty{10}{\celsius}. In order to measure both solid-like regime at the highest temperature and liquid-like regime at the lowest temperature, a 9 orders of magnitude experimental window is necessary. In practice, different techniques give access to different time windows.

\subsection{Rheometer}
Following \citet{Conrad2019} and \citet{Palombo2025}, we characterize the low temperature mechanical response of our system using bulk oscillatory rheology. This technique consists in applying a small amplitude oscillatory shear deformation $\gamma(t) = |\gamma|\cos(\omega t)$ on a macroscopic sample (here \qty{70}{\micro\litre}) and measure the resulting oscillatory shear stress $\sigma(t)=|\sigma|\cos(\omega t+\delta)$, where $\omega$ is the angular frequency and $\delta$ the loss angle. In complex notations, $\sigma = G^*\gamma$, where $G^*=|G^*|e^{\imath\delta} = G^\prime +\imath G^{\prime\prime}$ is the complex shear modulus, which real and imaginary parts are respectively the storage modulus and the loss modulus. $\tan\delta = G^{\prime\prime}/G^\prime$ is called the loss tangent. The frequency dependence $G^*(\omega)$ is our mechanical response function. To construct it several measurements are done at different frequencies spaced logarithmically to sample the frequency range.

The frequency range of a rheometer is limited by inertia at high frequency and by torque resolution of diverging experimental time at low frequency. In the case of our geometry (cone-plate, with a metallic cone of radius \qty{25}{\mm} and an angle $\alpha = \qty{1}{\degree}$ truncated at \qty{50}{\micro\m}), we cannot reach beyond $\omega_\mathrm{max}=\qty{100}{\radian/\second}$, i.e. we are limited to time scales longer than $\approx\qty{0.1}{\second}$. On the low frequency end, we are limited by the sensitivity of the rheometer. Indeed the viscous response scales as $\omega$, thus the stress becomes very small at low frequency. In practice, our lowest frequency is $\omega_\mathrm{min}=\qty{e-3}{\radian/\second}$, where a single point of the frequency spectrum is acquired in more than \qty{1}{\hour}.

We use a stress-controlled rheometer MCR 301e from Anton Paar with a Peltier temperature controlled bottom plate. To ensure sample thermal equilibration at the set temperatures, our maximum cooling rate is \qty{1}{\kelvin/\minute}. The three solutions, each containing one of the DNA strands are mixed directly onto the plane of the rheometer at \qty{20}{\celsius}. After the lowering of the rotor, a ring of sunflower oil is added around the cone, and the whole is covered  with a solvent trap to minimize sample dry out. The sample is brought to \qty{90}{\celsius} and then gently mixed by rotating the cone. Then it is annealed, before measuring its properties at the wanted temperature. To ensure that no oil is mixed with the sample, we systematically conducted frequency sweeps at a reference temperature (at \qty{20}{\celsius} or \qty{25}{\celsius}, depending on the design) between successive tests, notably pre-/post-annealing. If the results were inconsistent, it was regarded as sample degradation and the sample was discarded. We have also performed tests without an oil ring, that were comparable but where limited in time due to evaporation.

The resolution of bulk oscillatory rheology is limited by the resolution in torque of the rheometer, $\mathcal{T}_\mathrm{min}\approx\qty{0.1}{\micro\newton\metre}$ in our case. We obtain the torque $\mathcal{T}$ for each point and use it to define the relative error of each quantity: $\mathrm{Error}[G^\prime]/G^\prime=\mathcal{T}_\mathrm{min}/\mathcal{T}$, same for the loss modulus $G^{\prime\prime}$, while $\mathrm{Error}[\tan\delta]/\tan\delta = 2\mathcal{T}_\mathrm{min}/\mathcal{T}$.

\subsection{DLS microrheology}
In a similar way as \citet{Fernandez2018}, we characterize the high temperature mechanical response of our system using dynamic light scattering (DLS) microrheology. This technique consists in measuring the spontaneous dynamics of microscopic tracer particles (\ce{PS-COOH} polystyrene coated with carboxyl groups of radius $a=\qty{255}{\nano\metre}$ from Bangs Laboratories) embedded at low concentration in the sample. The inertia of such small particles can be neglected at time scales larger than \qty{e-7}{\second}~\cite{makris2020ViscousviscoelasticCorrespondencePrinciple}. In a simple Newtonian fluid of viscosity $\eta_\mathrm{s}$, the motion of the particles should be diffusive, with a mean square displacement
\begin{equation}
    \langle\Delta r^2\rangle(\Delta t) = \frac{dk_\mathrm{B}T}{3\pi a}\frac{\Delta t}{\eta_\mathrm{s}},
\end{equation}
where $d=3$ is the dimension of space and $k_\mathrm{B}$ the Boltzmann constant. Therefore measuring $\langle\Delta r^2\rangle(t)$ and knowing the radius $a$ of the tracers allow the extraction of the viscosity. This Stokes-Einstein relation can be generalized to viscoelastic materials as~\cite{Mason1995}
\begin{equation}
    \langle\Delta r^2\rangle(\Delta t) = \frac{dk_\mathrm{B}T}{3\pi a}J(\Delta t),
    \label{eq:genStokesEinstein}
\end{equation}
where $J(\Delta t)$ is the same creep compliance as defined in section~\ref{sec:linearmech}. Eq.~(\ref{eq:genStokesEinstein}) means that particle dynamics $\langle\Delta r^2\rangle(\Delta t)$ is the response to thermal excitation $k_\mathrm{B}T$.

Using a commercial nanoparticle analyzer nanoPartica SZ-100V2 from Horiba, we can measure particle dynamics between \qty{e-6}{\second} and \qty{e-1}{\second}. The DNA strand and tracer solutions are mixed inside a \qty{70}{\micro\liter} quartz microcuvette (Hellma 105.251-QS.Z8,5), covered by mineral oil to prevent evaporation, and directly annealed in the temperature-controlled DLS setup. We use a holder adapter to ensure good thermal conduction. After annealing, temperature is decreased by steps of $\qty{1}{\kelvin}$, equilibrated for \qty{5}{\minute}, and then 5 acquisitions of \qty{2}{\minute} each are performed. We perform 5 coolings on each sample.

Each acquisition consists in shining a \qty{532}{\nano\m} laser through the cuvette and collecting the scattered light $I(q, t)$ at a \qty{90}{\degree} angle to a photomultiplier, yielding the autocorrelation function of the scattered intensity
\begin{equation}
    g_2(q,\Delta t) -1 = \frac{\langle I(q, t)I(q, t+\Delta t)\rangle_t}{\langle I(q, t)\rangle_t^2},
\end{equation}
where $q\approx\qty{23}{\per\micro\metre}$ is the wavenumber and $\langle\ldots\rangle_t$ denotes time average. To ensure that the time average is equivalent to an ensemble average (ergodic condition, see Ref\cite{pusey1989DynamicLightScattering, cipelletti:hal-04854424}), we keep only acquisitions where $2-g_2(q, \Delta t_\mathrm{min})<\num{e-2}$. 
At each temperature the remaining acquisitions are averaged into $\bar{g}_2(q, \Delta t)$ and converted to intermediate scattering function $f(q,\Delta t) = \sqrt{\bar{g}_2(q,\Delta t)-1}$. If at a given temperature $f(q,\Delta t_\mathrm{max})>0.1$, we consider that the system is not ergodic and we discard this temperature and all temperatures below. Ultimately, we obtain the creep compliance as
\begin{equation}
    J(\Delta t) = -\frac{2d\pi a}{q^2 k_\mathrm{B}T}\log f(q,\Delta t).
    \label{eq:JfromISF}
\end{equation}

We consider two sources of error on $g_2$: numerical errors that are important at short times where $2-g_2\leq 1$, and statistics of time averaging that are important at long $\Delta t$. We set the upper bound of the former to $2-\bar{g}_2(q,\Delta t_\mathrm{min})$, whereas the latter has been estimated to $6\sqrt{\Delta t/t_\mathrm{avg}}$, where $t_\mathrm{avg}=\qty{2}{\minute}$ is the total averaging time~\cite{degiorgio1971IntensityCorrelationSpectroscopy, Fernandez2018}. The error on the intermediate scattering function is then
\begin{equation}
    \mathrm{Error}\left[f(q,\Delta t)\right] \approx 1-f(q,\Delta t_\mathrm{min}) + \frac{3}{f(q,\Delta t)}\sqrt{\frac{\Delta t}{t_\mathrm{avg}}}.
\end{equation}
Errors bars on the creep compliance are then obtained by carrying explicitly $f\pm\mathrm{Error}\left[f\right]$ through the very non linear Eq.~(\ref{eq:JfromISF}). Points where relative error on $J$ is larger than 1 are discarded.

For the ergodic condition to be satisfied, we need at least two decades of liquid-like behaviour within our time window. That sets the maximum characteristic time we can measure with our DLS apparatus to \qty{e-3}{\second}.

\subsection{Methods we will not use}
It is possible to perform DLS on DNA samples without tracers~\cite{Biffi2013,Biffi2015,Nava2017}. In this case, the technique probes the fluctuations of the DNA molecules, e.g. the relaxation of the nanostar network. In the absence of well-defined tracers it is not possible to deduce mechanical response function, and only time scales can be deduced, neither modulus nor viscosity.

The dynamics of tracers can also be obtained from experiments under a microscope~\cite{Mason1997,ederaDifferentialDynamicMicroscopy2017}. The accessible time scales are then limited by the frame rate of the camera, seldom above \qty{1000}{fps}. Furthermore, the spatial resolution is not as good as DLS. Obtaining displacements below \qty{30}{\nano\meter} is difficult, which makes moduli larger than a few \unit{\pascal} impossible to measure. These techniques are thus often limited to the quantification of fluid regimes~\cite{Fernandez2018}.

Finally, Dynamic Wave Spectroscopy (DWS) is a variant of DLS based on the multiple scattering of the light. The ergodic condition can be satisfied much more easily in DWS, which greatly extends the capacity to probe slowly relaxing mechanical responses. DWS has been used for microrheology of DNA hydrogels~\cite{Xing2018} but was not accessible to us at the time of this study.

\section{Results and discussion}
\subsection{Low temperature results}

\begin{figure}
    \includegraphics[width=\linewidth]{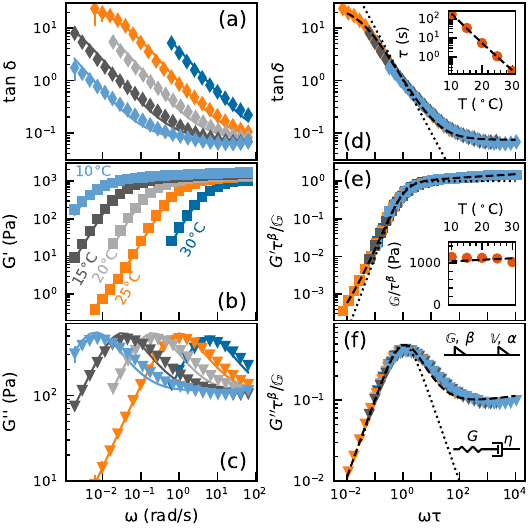}
    \caption{Bulk rheology of \texttt{Y16SE6} at \qty{1}{\milli\molar} at temperatures below $T_\mathrm{SE}$. (a) Loss factor (b) storage modulus, (c) loss modulus function of the frequency. Solid lines are fits of the fractional Maxwell model (fMM). 
    (d-f) Time-temperature superposition of the previous. Black dashed lines show the behavior of the fMM for $\alpha=0.98$ and $\beta=0.05$, top diagram in (f). Black dotted lines show the behaviour of the non fractional Maxwell model, bottom diagram in (f). 
    Insets of (d) and (e) present the temperature dependence of the fMM fit parameters, respectively the crossover characteristic timescale and the characteristic elasticity. Black dashed lines show respectively Arrhenius fit of activation energy \qty{241}{\kilo\joule/\mol} and modulus prediction from nanostar concentration.
    }
    \label{fig:tts}
\end{figure}

Figure \ref{fig:tts}(a-c) shows the results of the frequency sweeps conducted at different temperatures of \texttt{Y16SE6} at $C_\mathrm{NS}=\qty{1}{\milli\molar}$. For all temperatures we observe $\tan\delta>1$ at low frequencies, that is $G^\prime < G^{\prime\prime}$, corresponding to a fluid-like behaviour. The loss tangent decreases with increasing frequency, reaching solid-like behaviour.
Consistently, the storage modulus $G^\prime$ increases at low frequencies until it reaches a plateau at high frequencies; whereas the loss modulus $G^{\prime\prime}$ increases more sharply than $G^\prime$ at low frequencies then reaches a maximum before it starts decreasing at high frequencies but soon reaches a plateau.

On Figure~\ref{fig:tts}(d-f) we show the same data shifted by temperature dependent factors so that they collapse on a master curve. This procedure, called `time-temperature superposition' is performed in two steps. 
(i) The frequency axis is scaled by a time $\tau(T)$ so that $\tan\delta(\omega\tau)=1$. The resulting collapse is shown on Figure~\ref{fig:tts}(d). 
(ii) The modulus axis is scaled by a common factor for both $G^\prime$ and $G^{\prime\prime}$. The resulting collapses are shown on Figure~\ref{fig:tts}(e,f).

On Figure~\ref{fig:tts}(d) we observe a frequency dependence of $\tan\delta$ that is more complex than the $1/\omega$ dependence expected for a Maxwell model (dotted line). The two plateaus at low and high frequencies and the slightly decreased slope are better described by a \emph{fractional} Maxwell model~\cite{Jaishankar2013} (dashed line). As the classical Maxwell model, the fractional Maxwell model consists in two mechanical elements in series. However these elements are spring-pots of respective exponents $\alpha>\beta$ and respective quasi-properties $\mathbb{V}$ and $\mathbb{G}$. The good collapse on a single master curve indicates that $\alpha=0.98$ and $\beta=0.05$ can be considered constant across the temperature range.
In terms of storage modulus (Fig.~\ref{fig:tts}e), the difference between classical and fractional Maxwell is significant only at the extremities of the frequency range. By contrast the loss modulus (Fig.~\ref{fig:tts}f) is well described only by the fractional model.
The fractional description suggests a broad distribution of relaxation processes. In particular at times much shorter than the main relaxation time, when the network of SE-SE bonds can be considered permanent, relaxations already occur.
As noted above, different pairs of arms have different binding free energies and thus different life times. 
Furthermore, we cannot exclude stoichiometry mismatch between the three DNA oligomers, causing a small population of nanostars lacking one or more oligomers~\cite{chaderjian2026DiverseDistinctDensely}.
In addition, synthesis errors shorten $P(n<N|n\geq N-n_\mathrm{SE})\approx 5.8\%$ of the sticky ends belonging to well-formed nanostars, leading to an additional handful of possible shorter life times.
Nevertheless, the distribution of bond lifetimes is quantified, whereas the power-law behaviour implies a continuous distribution of relaxation times.
We think that the intrinsically disordered structure of nanostar gels~\citet{Rovigatti2014} can explain the transition from quantified to continuous distribution.
Indeed, at the macroscopic scale of our measurements, the variety of possible spatial arrangements of bonds of different relative lifetimes is practically infinite, leading to a continuous population of possible environments relaxing at different rates.

The inset of Figure~\ref{fig:tts}(d) shows the temperature dependence of $\tau$. As expected, it follows an Arrhenius dependence compatible with Eq.~(\ref{eq:tau}) with $n\approx 1.14$. This indicates that rearrangement involves a single bond dissociation, consistently with~\cite{Conrad2019}.
The inset of Figure~\ref{fig:tts}(e) shows the temperature dependence of $\mathbb{G}/\tau^\beta$, i.e. the factor that allows the collapse of the moduli. It corresponds to the value that would have the quasi-elastic plateau at $\Delta t =\tau$. The values are compatible with the simple scaling $G=\rho_\mathrm{NS}k_\mathrm{B}T/2$, where $\rho_\mathrm{NS}$ is the number density of nanostars~\cite{Nava2018, Palombo2025}. This is consistent with the picture of a dense network where every nanostar is a cross-link that freezes in a microscopic degree of freedom.

We have seen that for $T<T_\mathrm{SE}$ the mechanical behaviour of the DNA nanostar gel is better described by a fractional Maxwell model that implies collective relaxation processes, in particular in the solid-like high frequency regime. Besides this new physics, however, most of the mechanical behaviour is still understandable from the picture of a percolated network, which dynamics is governed by the lifetime of the SE-SE bond. Within this picture, no elasticity is expected above $T_\mathrm{SE}$ where the network is not percolated~\cite{Nava2018}.


\subsection{High temperature results}

\begin{figure}
    \includegraphics[width=\linewidth]{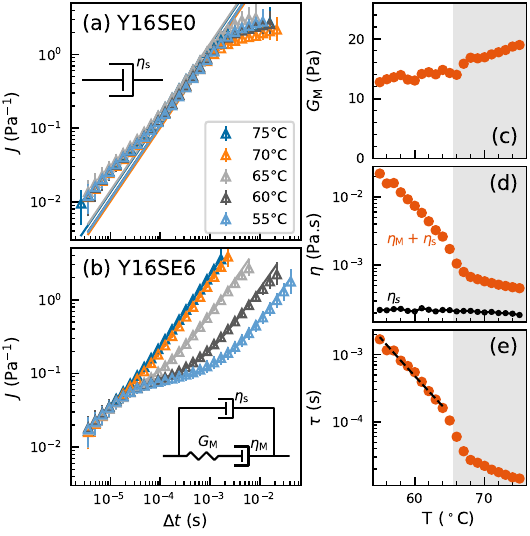}
    \caption{DLS microrheology in \texttt{Y16SE0} vs. \texttt{Y16SE6} at \qty{1}{\milli\molar} for temperatures above $T_\mathrm{SE}$. 
    (a) Compliance of nanostars without sticky ends. Solid lines are Newtonian fits. 
    (b) Compliance of nanostars with sticky ends. Solid lines are Jeffreys model fits. 
    (c,d,e) Temperature dependence of the parameters of Jeffreys fits of \texttt{Y16SE6}: (c) elasticity modulus $G_\mathrm{M}$, (d) viscosity $\eta_M$ and (e) characteristic time $\tau=\eta_\mathrm{M}/G_\mathrm{M}$. Gray area indicates the fluid temperature range where $\tan\delta>1$ at all times. Dotted black line on (e) is an Arrhenius law of same activation energy as in the inset of Fig.~\ref{fig:tts}(d).}
    \label{fig:dls-results}
\end{figure}

Figure~\ref{fig:dls-results}(a) shows the compliance of a suspension of nanostars without sticky-ends (\texttt{Y16SE0}) for several temperatures between $T_\mathrm{NS}$ and $T_\mathrm{SE}$, obtained by DLS microrheology. 
All temperatures are superimposed and, in the time range where error bars are small, the time dependence is linear. This corresponds to the behaviour of a Newtonian liquid of viscosity $\eta$ that varies little in this temperature range. This is the expected behaviour in the absence of link between nanostars.

By contrast, for nanostars with sticky-ends (\texttt{Y16SE6}) the shape of the compliance curve evolved strongly with temperature, as shown on Figure~\ref{fig:dls-results}(b). From a Newtonian behaviour at the highest temperature, we observe the development of a plateau at intermediate times as $T$ is decreased. At the lowest temperature accessible to DLS microrheology, that is still $\approx\qty{10}{\kelvin}$ above $T_\mathrm{SE}$, the plateau spans almost 2 orders of magnitude in time. The presence of such a plateau is the unmistakable proof of an elastic regime. At shorter times, the time dependence is mostly linear with a corresponding viscosity close to the one of the buffer. After the plateau, the time dependence becomes linear again but with a much larger viscosity. Such a behaviour is also reminiscent of the behaviour of a supercooled liquid close to the glass transition~\cite{Cavagna2009}.

In order to describe quantitatively these three mechanical regimes, we use the Jeffreys model, that is a viscosity $\eta_\mathrm{s}$ in parallel to a Maxwell model of modulus $G_\mathrm{M}$ and viscosity $\eta_\mathrm{M}$. This model has two time scales: the entry to the plateau at $\tau_\mathrm{s}=\eta_\mathrm{s}/G_\mathrm{M}$, and the exit of the plateau at $\tau=\eta_\mathrm{M}/G_\mathrm{M}$. Conceptually, $\eta_\mathrm{s}$ is the viscosity of the solvent (the buffer) in which nanostars are suspended, whereas $G_\mathrm{M}$ and $\eta_\mathrm{M}$ characterize the contribution of DNA nanostars. Here, we set aside on purpose the possibility of fractional elements in order to keep a small number of parameters. Despite this parsimony, the analytical expression of the compliance of the Jeffreys model fits well the experimental data at all accessible temperatures
. The fitted modulus, shown in Fig.~\ref{fig:dls-results}(c) is mostly constant and at least two orders of magnitude lower than the modulus measured in bulk rheology (inset of Fig~\ref{fig:tts}e). Viscosities are shown in Fig~\ref{fig:dls-results}(d). The solvent viscosity $\eta_\mathrm{s}$ remains mostly constant, following the temperature dependence of water (not shown). By contrast, the long term viscosity $\eta_\mathrm{M}+\eta_\mathrm{s}$ increases by two orders of magnitude in the accessible temperature range. This increase roughly mirrors the increase of the relaxation time of the network $\tau=\eta_\mathrm{M}/G_\mathrm{M}$ shown in Fig.~\ref{fig:dls-results}(e). Below \qty{65}{\celsius}, $\tau(T)$ is also well described by an Arrhenius law of same activation energy as found from bulk rheology below $T_\mathrm{SE}$ (inset of Fig~\ref{fig:tts}d). We are thus confident that the dissociation of the SE-SE bonds is still setting the relaxation time of the system.

\begin{figure*}
    \centering
    \includegraphics[width=\linewidth]{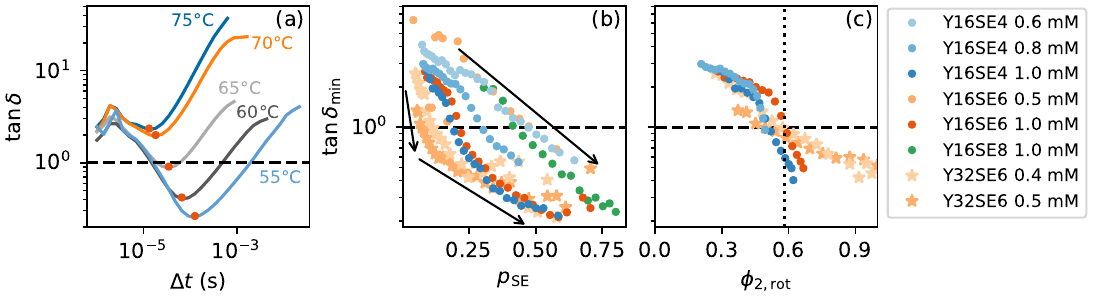}
    \caption{Effect of sequence design and concentration on the onset of solidity. 
    (a) For Y16SE6 at \qty{1}{\milli\molar}, the loss tangent from DLS microrheology function of delay time for several temperatures. Orange circles highlight the minimum at each temperature. Below the dashed line, the behavior is solid-like.
    (b) For several designs and concentrations, the value of the minimum of $\tan\delta$ at temperature $T$ function of the probability of SE-SE bond at the same temperature. Arrows highlight the two extreme behaviours upon temperature decrease.
    (c) Same as (b) but function of the volume occupied by freely rotating doublets at the same temperature. The vertical dotted line is at $\phi_{2,\mathrm{rot}}=0.58$. For clarity, we plot only designs and concentrations where percolation cannot explain solidity, and temperatures where doublets are the majority clusters.
    }
    \label{fig:all_designs}
\end{figure*}

Modeling the mechanical response by the Jeffreys model allows to quantify well-defined material properties. However, even with a non zero internal modulus $G_\mathrm{M}$, Jeffreys model can display a response that is mostly fluid at all time scales if $\tau$ is not much larger than $\tau_s$. In other words, if the plateau is too short the curve $J(\Delta t)$ will not have a flat portion. Around a time scale $\Delta t$, one can approximate the compliance by a power law of exponent $\alpha = (\mathrm{d}\log J)/(\mathrm{d}\log\Delta t)$~\cite{Mason1995}. As for the spring-pot element, $\alpha=0$ indicates a purely elastic response at this time scale, $\alpha=1$ a purely viscous response. More generally, the local exponent is linked to the loss tangent $\tan\delta = \tan\left(\frac{\pi}{2}\alpha\right)$. In particular $\alpha<0.5\Leftrightarrow\tan\delta<1$ means that the solid response dominates at this time scale. In Figure~\ref{fig:all_designs}(a) we display the time dependence of the loss tangent obtained from the local power law approximation of $J(t)$. We observe that for $T>\qty{65}{\celsius}$ we have $\tan\delta>1$ at all times, implying a dominant fluid response. However at lower temperatures, the material exhibits a dominantly solid response at least at some time scales. We thus confirm the emergence of elasticity some \qty{10}{\kelvin} above $T_\mathrm{SE}$.

The emergence of a short-time elasticity well above the melting temperature of the SE-SE bonds is the main result of this manuscript and cannot be explained by arguments based on percolation of the DNA nanostar network. Indeed, \texttt{
Nupack} calculations predict that between \qty{60}{\celsius} and \qty{65}{\celsius}, the probability $p_\mathrm{SE}$ for a full-length SE-SE bond to be associated ranges from 0.33 to 0.22, well below percolation threshold at $p_\mathrm{SE}=0.5$. These probabilities are further reduced by $(1-q)^{n_\mathrm{SE}}$ due to synthesis error causing shorter sticky ends. At these values of $p_\mathrm{SE}$, there is no percolating cluster to transmit stress. Furthermore, one can compute the probability for a nanostar to form a single bond, respectively two bonds as $P(b=1) = 3p_\mathrm{SE}(1-p_\mathrm{SE})^2$ and $P(b=2) = 3p_\mathrm{SE}^2(1-p_\mathrm{SE})$.
When elasticity emerges at \qty{65}{\celsius}, $P(b=1)\approx0.39$ and $P(b=2)\approx0.10$, meaning that most clusters of nanostars are doublets. More generally, as long as $P(b=2)<P(b=1)/2\Leftrightarrow p_\mathrm{SE}<1/3$, doublets are the majority clusters and the concentration of doublets can be approximated as $C_2 \approx C_\mathrm{NS}P(b=1)/2$.

In the absence of percolation of the DNA network, we look for a different cause of elasticity: glass transition. On Figure~\ref{fig:phase-diag}, we observe that the packing fraction of nanostars is below $0.45$ at our concentration, thus below usual thresholds for glass transition around 0.56-0.58. This packing fraction is computed by considering that each nanostar occupies the volume in which it is free to rotate.
However, if we now consider a doublet of nanostars, its free rotation requires a volume much larger than twice the volume of an individual nanostar. Indeed, the total contour length of a doublet is $L=(4n_\mathrm{Y}+n_\mathrm{SE})u+4r$. Its effective length $L_\mathrm{eff}$ is obtained from Eq.~(\ref{eq:crossoverSAW}). The Brownian rotational time of a doublet of \texttt{Y16SE6} is $\tau_\mathrm{rot,2}\approx\qty{7}{\micro\second}$. Therefore when DLS measurements indicate an elastic plateau, a doublet is randomly rotating and occupies a volume $V_\mathrm{rot,2} = \pi L_\mathrm{eff}^3/6$.
The effective volume fraction of doublets is thus $\phi_\mathrm{rot,2} = V_\mathrm{rot,2}C_2\mathcal{N}_\mathrm{A}$. In our case, at $T=\qty{65}{\celsius}$ where we observe the emergence of solidity, we find that $\phi_\mathrm{rot,2}\approx 0.56$, which corresponds to a usual threshold for the glass transition.

To go beyond a mere coincidence, we repeat our DLS microrheology experiments at various nanostar concentrations. We also alter the nanostar design by varying the length of the sticky ends and the length of the arms. In order to monitor systematically the emergence of solidity in these samples, we repeat the method displayed in Figure.~\ref{fig:all_designs}(a): at each temperature we compute the loss tangent, we pick up the minimum value (orange circles) and thus obtain $\tan\delta_\mathrm{min}(T)$. Solidity emerges when $\tan\delta_\mathrm{min}<1$. On Figure~\ref{fig:all_designs}(b), we display $\tan\delta_\mathrm{min}(T)$ function of the SE-SE binding probability $p_\mathrm{SE}(T)$ (obtained from \texttt{Nupack}) for all investigated designs and concentrations. 
We observe a range of behaviours between two limiting cases.
For long sticky ends (\texttt{Y16SE8} at \qty{1}{\milli\molar}) or low concentrations (\texttt{Y16SE6} at \qty{0.5}{\milli\molar} and \texttt{Y16SE4} at \qty{0.6}{\milli\molar}), we observe an almost exponential decay of $\tan\delta_\mathrm{min}$ with respect to $p_\mathrm{SE}$. Solidity is reached near $p_\mathrm{SE}=0.5$, which is consistent with the percolation of the nanostar network and previous studies~\cite{Nava2018}.
By contrast, for long arms (\texttt{Y32SE6}) and high concentrations  (\texttt{Y16SE4} and \texttt{Y16SE6} at \qty{1}{\milli\molar}), the exponential regime is preceded by an initial drop in $\tan\delta_\mathrm{min}$ that reaches values below 1 while $p_\mathrm{SE}$ increases little.
This means that $p_\mathrm{SE}$ alone fails to describe the emergence of elasticity. We thus confirm that in this last case network percolation is not the important physical ingredient behind the emergence of elasticity in our system.


On figure~\ref{fig:all_designs}(c), we display $\tan\delta_\mathrm{min}(T)$ function of the packing fraction of freely rotating doublets $\phi_\mathrm{rot,2}(T)$ for the samples where percolation cannot explain the emergence of solidity. For clarity, we plot only temperatures where $p_\mathrm{SE}<1/3$ so that our estimate of the concentration of doublets is correct.
We observe that for all remaining samples solidity is reached for $\phi_\mathrm{rot,2}\approx 0.55-0.6$, despite differences in binding probability, concentration and arm length.
Our observations underline the relevance of the volume occupied by rotating doublets. When the doublets crowd each-other their rotational degrees of freedom freeze and the whole system acquire a short-time elasticity of glassy origin. At time scales longer than the life time of the SE-SE bonds, doublets are transient and a fluid behaviour is recovered.

\section*{Conclusions}

We have investigated the mechanical response of a concentrated suspension of DNA nanostars of simple design, both at low temperatures by bulk oscillatory rheology, and at high temperatures by dynamic light scattering microrheology. 

Below the melting temperature of the sticky-ends, we have evidenced by time-temperature superposition over 6 orders of magnitude in frequency a mechanical response well-described by a \emph{fractional} Maxwell model. This description is more precise than previous `Maxwellian` typology, and highlights already observed -- but not interpreted -- differences with the canonical Maxwell model~\cite{Conrad2019}. In particular the fractional Maxwell model quantifies the quasi-elastic behaviour at high frequency, i.e. the slope of the plateau of $G^\prime$, suggesting a broad range of relaxation times that stem both from a distribution in bond lifetimes and from the heterogeneous disordered microstructure of the gel. We have thus demonstrated that a macroscopic mechanical test is able to quantify the disorder of the DNA assembly at the nanoscale.

Above the melting temperature, we have reported an unexpected solid behaviour at times shorter than the millisecond, responsible for a two order of magnitude increase in the viscosity felt at more usual time scales. A similar viscosity increase was reported by \citet{Fernandez2018} who assigned it to the progressive formation of the viscoelastic network. Here, we have demonstrated that the emergence of solidity is not always linked to the percolation of the nanostar network, expected \qty{10}{\kelvin} lower~\cite{Nava2018}. Instead, we have built a model based on the idea that the crowding of doublets of nanostars hinders their rotation, effectively locking degrees of freedom into a glass transition that induces material solidity. We have verified this model experimentally by varying the nanostar concentration and design. This model goes beyond the description of DNA nanostars as strong glass formers~\cite{Biffi2015}, and is more akin to the glass transition of dumbbells~\cite{chong2002IdealizedGlassTransitions} or other anisotropic particles~\cite{kramb2010YieldingDenseSuspensions}. We have thus shown that at high enough packing fraction the solidity of DNA nanostar suspensions emerge in two steps: first a glass transition due to the crowding of rotating doublets, and second the percolation of a load-bearing DNA network.

We have also presented a methodology for mechanical investigation of DNA materials, from identifying the regions of the phase diagram where mechanical measurements are possible, to selecting measurement methods to cover the time scales, to the measurements of mechanical response functions and their modeling in terms of material properties, to hypotheses on nanoscale processes, to iterations on the design to validate the hypotheses. We hope that this method will help the nanoscale design community in embracing tools from mechanical measurement, soft matter and statistical physics to characterize and understand disordered DNA materials.

\section*{Author contributions}
Investigation, including all experiments, were performed by H.A. 
Conceptualization and supervision by A.G., N.S., H.G., C.B. and M.L.
Resources and funding acquisition provided by A.G., S.H.K., H.G., C.B and M.L.
Software provided by N.S. and M.L.
Data curation, formal analysis, validation, visualisation and original draft writing by H.A. and M.L.
Methodology and project administration by M.L.
Review and editing by H.A. C.B. and M.L.

\section*{Conflicts of interest}
There are no conflicts to declare.

\section*{Data availability}
Data for this article, including results of \texttt{Nupack} and \texttt{UNAfold} simulations, bulk rheology measurements, DLS measurements, as well as the \texttt{Matplotlib}~\cite{Hunter:2007} scripts to generate the figures are available at Zenodo at \url{https://doi.org/10.5281/zenodo.19133765}


\section*{Acknowledgements}
H.A. acknowledges PhD funding from CNRS Prime program and postdoctoral funding from FY2024 JSPS Postdoctoral Fellowship for Research in Japan (short-term).
H.A. and M.L. acknowledge funding from ANR ActivSoliPolar (ANR-24-CE30-6841).
H.A., H.G., S.H.K. and M.L. acknowledge funding from LIMMS Internal projects.
M.L. acknowledges funding from ANR VitriPSA (ANR-21-CE06-0025-02) and CNRS IRP RePAS.
S.H.K. acknowledges funding from ANR DEVILISH (ANR-24-CE50-4784).
H.G. acknowledges funding from CNRS MITI ElectroDSC.
H.A. and M.L thank Jan Eichler and Bruno Issenman for guidance on DDM experiments that were unfortunately not exploitable.
H.A., A.G. and S.H.K. thank Kazutoshi Masuda and Miho Yanagisawa for guidance on micropipette experiments that were unfortunately not exploitable.
The authors thank Luca Cipelletti for fruitful discussions and catching some of our misinterpretations of the DLS data. All remaining mistakes are ours.

This article is dedicated to the memory of Anthony Genot.



\balance


\bibliography{biblio} 
\bibliographystyle{rsc} 


\end{document}